\documentclass[aip, jcp,  superscripts, reprint]{revtex4-1}
\usepackage{amstext}
\usepackage{amsmath}
\usepackage{amssymb}
\usepackage{graphicx}
\usepackage{psfrag}
\usepackage{tensor}

\usepackage[usenames,dvipsnames]{color}

\newcommand{\eq}[1]{Eq.~\eqref{#1}}
\newcommand{\fig}[1]{Fig.~\ref{#1}}
\newcommand{\stn}[1]{Sec.~\ref{#1}}
\newcommand{\be}{\begin{equation}}
\newcommand{\ee}{\end{equation}}

\newcommand{\mc}[1]{\mathcal{#1}}

\newcommand{\ket}[1]{| #1 \rangle}
\newcommand{\bra}[1]{\langle #1 |}

\newcommand{\w}{\omega}

\begin{document}
\title{Reduced purities as measures of decoherence in many-electron systems}
\author{Ignacio Franco}
\altaffiliation[Current address: ]{Department of Chemistry, University of Rochester, Rochester, New York, 14627 USA }
\email{franco@chem.rochester.edu}
\affiliation{Theory Department, Fritz Haber Institute of the Max Planck Society, Faradayweg 4-6, 14195 Berlin, Germany}

\author{Heiko Appel}
\affiliation{Theory Department, Fritz Haber Institute of the Max Planck Society, Faradayweg 4-6, 14195 Berlin, Germany}

\date{\today}

\begin{abstract}
I. Franco and H. Appel J. Chem. Phys. \textbf{139}, 094109 (2013) \\ \\
A hierarchy of  measures of decoherence for many-electron systems that is based on the purity and the hierarchy of reduced electronic density matrices is presented. 
These reduced purities can be used to characterize electronic decoherence  in the common case when the many-body electronic density matrix is not known and only reduced information about the electronic subsystem is available. Being defined from reduced electronic quantities, the interpretation of the reduced purities is more intricate than the usual (many-body) purity. This is because the nonidempotency of the $r$-body  reduced electronic density matrix that is the basis of the reduced purity measures can arise due to decoherence or due to electronic correlations. To guide the interpretation, explicit expressions are provided for the one-body and two-body reduced purities for a general electronic state. Using them, the information content and structure of the one-body and two-body reduced purities is established, and limits on the changes that decoherence can induce are elucidated. The practical use of the reduced purities to understand decoherence dynamics in many-electron systems is exemplified through an analysis of the electronic decoherence dynamics  in a model molecular system.   
\end{abstract}

\maketitle

\section{Introduction}

An ubiquitous process in nature is that of decoherence~\cite{breuer, schlosshauer, joos, zurek}, which refers to the decay of quantum correlations of a quantum subsystem because of interaction with an environment. Understanding decoherence is central to our description of basic processes such as measurement, photosynthesis, vision or electron transfer~\cite{martinez2007, fleming09, hannewald09, choi2010, engel07,collini09, brumer_2012, pachon_2012, pachon_2013}, to the development of approximation schemes that describe the system-bath dynamics~\cite{ray, breuer, weiss, winzer2010, malic2011, yuen2009, yuen2010, heiko09, heiko2011} and it is the starting point for the design of methods to preserve coherence in materials that can be subsequently exploited in intriguing and potentially useful ways via quantum control~\cite{pbnewbook, ricebook} or quantum information~\cite{nielsen}  schemes.

Here we introduce measures of decoherence that can be used  for the description of the coherence properties of many-electron systems in the presence of an environment.  The proposed measures are generalizations of the purity that employ the few-body electronic reduced density matrices instead of the full many-body reduced density matrix,  and are thus more readily available for the characterization of coherence in many-body systems. However, because these measures are based on reduced electronic quantities their interpretation differs in key aspects and can be more intricate from the one of the usual $N$-body purity. Below we discuss the properties, merits and limitations of the reduced purity measures. 

The structure of this manuscript is as follows.  In \stn{stn:purity} we briefly review basic decoherence ideas as they apply to many-electron systems. Then, in \stn{stn:reducedpurities}, we introduce a hierarchy of reduced purity measures that are based on the well-known hierarchy~\cite{cohen1976, bonitz} of reduced electronic density matrices. In particular, we determine analytical expressions for the reduced purities  that follow from the one- and two-body reduced density matrices for a general time-dependent correlated electronic state (Secs.~\ref{stn:slater}-\ref{stn:p1andp2}). Using these expressions,  we then discuss in \stn{stn:interpretation} the interpretation of the reduced purities and the effect of electronic correlation on their evolution. Last, in \stn{stn:examples} we exemplify the use of the reduced purities by studying electronic decoherence in a vibronic system. We summarize our main findings  in \stn{stn:conclusions}.

\section{Purity and the interpretation of  decoherence}
\label{stn:purity}

Consider an $N$-particle electronic system interacting with an environmental bath, with system-bath Hamiltonian of the form
$H = H_e + H_B + H_{e-B}$,
where $H_e$ is the electronic Hamiltonian, $H_B$ the bath Hamiltonian and $H_{e-B}$ is the system-bath coupling.  In light of the Schmidt decomposition~\cite{nielsen}, a pure state of the bipartite system can always be written as an entangled state of the form
\be
\label{eq:schmidt}
\ket{\Omega (t)} = \sum_n \sqrt{\lambda_n} | \Psi_n \rangle | B_n \rangle,
\ee
where $\ket{\Psi_n}$  are orthonormal states of the electronic subsystem  and $| B_n \rangle$ orthonormal states of the bath. The Schmidt coefficients $\sqrt{\lambda_n}$ are nonnegative real numbers satisfying $\sum_n \lambda_n = 1$. It is often useful to express \eq{eq:schmidt} in terms of the $N$-particle eigenbasis $\ket{E_n}$ of the many-electron Hamiltonian $H_e$. Since the $\{\ket{E_n}\}$ form a complete set in the subsystem Hilbert space, in general
$\ket{\Psi_n} = \sum_m {c}_{mn} \ket{E_m}$. Thus, \eq{eq:schmidt} can be rewritten as
\be
\label{eq:schmidt2}
\ket{\Omega (t)} = \sum_n \ket{E_n} \ket{\chi_n(t)},
\ee
where the bath states  associated with each of the $\ket{E_n}$  are defined by
$ \ket{\chi_n} =  \sum_m {c}_{nm} \sqrt{\lambda_n} \ket{B_m}$.
 The $\{ \ket{\chi_n}\}$ are not orthonormal but do satisfy $\sum_n \bra{\chi_n}{\chi_n}\rangle= 1$. 
 
The properties of the electronic subsystem for such an entangled state $\ket{\Omega}$ are completely characterized by the  $N$-particle electronic density matrix 
 \be
\label{eq:rrho}
\hat{\rho}_e(t) = \textrm{Tr}_B\{\ket{\Omega}\bra{\Omega}\}
=\sum_{nm} \langle \chi_m (t) | \chi_n (t) \rangle  | E_n\rangle \langle E_m |,
\ee
where the trace is over the environmental degrees of freedom. Note that the coherences or phase relationship between electronic eigenstates (the off-diagonal elements) in $\hat{\rho}_e$ are determined by the  overlaps $S_{nm}(t)=\langle \chi_m | \chi_n  \rangle $ between the environmental states associated with the electronic eigenstates. Thus, the loss of coherences in $\hat{\rho}_e(t)$ can be interpreted as the result of the decay of the $S_{nm}$ during the coupled electron-bath evolution~\cite{schwartzoleg_1996, hwang04, franco08, franco12}.  Standard measures of decoherence capture precisely this. For example, the purity, the measure of decoherence that we focus on here, is given by 
\be
\label{eq:purity}
P(t) = \textrm{Tr}\{\hat{\rho}_e^2(t)\} =  \sum_{nm} |\langle \chi_m(t)|\chi_n(t)\rangle |^2 = \sum_n \lambda_n^2
\ee
and decays with the overlaps between the environmental states $S_{nm}$.

\section{A hierarchy of reduced measures of electronic decoherence}
\label{stn:reducedpurities}

In order to quantify the coherence of a given many-particle electron system one ideally would like to study the $N$-body purity in \eq{eq:purity} directly.  It is simple to interpret ($P=1$ for pure states; $P<1$ for mixed states; $P= 1/M$ for a maximally entangled $M$-level subsystem), it has well defined upper and lower values and captures all possible electronic coherences. This, however, is not always possible because of the many-body nature of the problem. To determine the purity from a time-dependent simulation one has to either propagate the many-body electronic density matrix [\eq{eq:rrho}] or follow the dynamics of the bath. Either approach is intractable in general except for few-level problems because of the inherent difficulty in solving the many-body problem (see, e.g., Ref.~\onlinecite{fetter}) and/or because of the high-dimensionality of the objects involved.  A reduced method to capture the essential electronic coherences is thus desirable. 

Here we introduce a hierarchy of measures of coherence in many-particle systems that is based on the well-known hierarchy of many-particle reduced density matrices~\cite{cohen1976, bonitz}. Specifically, we define the $r$-body reduced purity (or $r$-body purity, for short) as
\be
\label{eq:purityeigen}
P_r(t) = \textrm{Tr} \{{}^{(r)}{\hat{\Gamma}}(t)^2\} = \sum_n {}^{(r)}\lambda_n^2,
\ee
where ${}^{(r)}{\hat{\Gamma}}$ refers to the $r$-body reduced electronic density matrix and the set $\{ {}^{(r)}\lambda_n\}$ to its eigenvalues.  The matrix elements of ${}^{(r)}{\hat{\Gamma}}$ can be expressed as
\be
\label{eq:rdm}
{}^{(r)}\Gamma^{j_1 j_2 \cdots j_r}_{i_1 i_2 \cdots i_r} (t) = \frac{1}{r!}\textrm{Tr}\{\hat{c}_{i_1}^\dagger \hat{c}_{i_2}^\dagger\cdots \hat{c}_{i_r}^\dagger \hat{c}_{j_r} \cdots \hat{c}_{j_2} \hat{c}_{j_1} \hat{\rho}_e (t) \}.
\ee
Here the  operator  $\hat{c}_{i}^\dagger$ (or $\hat{c}_{i}$) creates (or annihilates) a fermion in the $i$th  spin-orbital of the basis set  and satisfies the usual fermionic anticommutation rules ($\{\hat{c}_{i}, \hat{c}_{j}^\dagger\} = \delta_{i,j}$, $\{\hat{c}_{i}^\dagger, \hat{c}_{j}^\dagger\}=\{\hat{c}_{i}, \hat{c}_{j}\} = 0$). Note that from the $r$-body reduced density matrix one can obtain all other lower-order ($r-s$) density matrices by contractions of the indices, and thus all lower-order purities. In general,
\be
\sum_{i_p} {}^{(r+1)}\Gamma^{j_1\cdots i_p \cdots j_{r+1}}_{i_1\cdots i_p \cdots i_{r+1}} (t) = \frac{N-r}{r+1} {}^{(r)}\Gamma^{j_1 j_2 \cdots j_{r}}_{i_1 i_2 \cdots i_{r}} (t),
\ee
where $p$ is an integer between 1 and $r+1$. The fully contracted $r$-body density matrix yields
\be
\sum_{i_1, \cdots, i_r} {}^{(r)}\Gamma^{i_1\cdots i_r}_{i_1\cdots i_r} (t) =  \frac{N!}{r!(N-r)!}.
\ee
Using this notation, the $r$-body purity can be expressed as
\be
\label{eq:rpurity}
P_r (t) = \sum_{\substack{i_1, \cdots, i_r\\j_1, \cdots, j_r} }
{}^{(r)}{\Gamma}_{i_1, \cdots, i_r}^{j_1, \cdots, j_r}(t)
{}^{(r)}{\Gamma}_{j_1, \cdots, j_r}^{i_1, \cdots, i_r}(t). 
\ee
Note that, because the trace is independent of the basis, the expression above is valid in any complete single-particle basis-set. Further note that it is also possible to define the $r$-body purity based on the spin-contracted $r$-body electronic density matrix~\cite{francoNJP2013}. However, the structure of the spin-uncontracted version adopted here is simpler and more amenable to generalization.

Because the $r$-body purity in \eq{eq:purityeigen} is defined by the reduced density matrix ${}^{(r)}\hat{\Gamma}$ obtained by tracing over the bath \emph{and} $N-r$ electronic coordinates, it can be argued that the $P_r$ are a measure of decoherence due to interactions with the bath \emph{and} the traced out  electronic degrees of freedom. However, since electrons are indistinguishable there is no operator that can take advantage of electronic entanglements or distinguish between an electronic ``subsystem" and an electronic ``bath".  Thus, we view the reduced purities as measures  of the coherence of many-electron systems that can be used in the usual case when only partial information about the electronic system is known. Nevertheless, because of their reduced nature, their interpretation requires  more care and differs in a few key aspects from the $N$-body purity in \eq{eq:purity} in ways  that are discussed in detail below. 

\subsection{Using Slater determinants to define a coherence order}
\label{stn:slater}

We are concerned with the coherence properties of a general $N$-particle  correlated time-dependent electronic density matrix $\hat{\rho}_e$ [\eq{eq:rrho}]. Without loss of generality, it is   convenient to express  $\hat{\rho}_e$  in terms of Slater determinants as
\be
\label{eq:rho}
\hat{\rho}_e =  \sum_{n, m} a_{nm} \ket{\Phi_n}\bra{\Phi_m},
\ee
where  $\ket{\Phi_n}$ corresponds to a single Slater determinant with integer occupation numbers  in a given (arbitrary) single-particle basis $\ket{\epsilon} = \hat{c}_\epsilon^\dagger \ket{0}$, where $\ket{0}$ is the vacuum level. In writing \eq{eq:rho}, we have expanded the correlated electronic states in \eq{eq:rrho} in terms of a basis of Slater determinants,  i.e. in a full configuration interaction expansion $\ket{E_i}= \sum_{n} b_{ni}\ket{\Phi_n}$. The $a_{nn}$ in \eq{eq:rho} denote the population of Slater determinant  $n$, while the $a_{nm}$ refer to the coherences between the $n, m$ pair.   In this context, we define the \emph{order $s_{nm}$ of a given pair of Slater determinants}  $\ket{\Phi_n}$ and $\ket{\Phi_m}$ as the number of single particle transitions required to do a $\ket{\Phi_n} \to \ket{\Phi_m}$ transition. This quantity can be computed by
\be
\label{eq:order}
s_{nm} = N - \sum_\epsilon f_n(\epsilon)f_m(\epsilon),
\ee
where $f_n(\epsilon)$ is the distribution function of the single particle levels $\epsilon$ in state $\ket{\Phi_n}$. The quantity  $f_n(\epsilon)$ is defined by
\be 
\label{eq:distrib}
\bra{\Phi_n} \hat{c}_\epsilon^\dagger \hat{c}_{\epsilon'}\ket{\Phi_n}=
f_n(\epsilon) \delta_{\epsilon, \epsilon'},
\ee
and  takes values of 0 or 1 depending on whether the orbital level $\epsilon$ is occupied or not. The quantity $s_{nm}\in [0, N]$ and takes the value 1 for pairs of states that differ by single excitations, 2 for doubles, etc. \emph{We will refer to a coherence between states $\ket{\Phi_n}$ and $\ket{\Phi_m}$ as a coherence of order $s_{nm}$. }

We now use these definitions to discuss properties of the reduced purities.

\subsection{The $r$-body purity can only distinguish coherences of order $r$ or less}

First note that, \emph{because the $r$-body purity is constructed from the $r$-body density matrix, it is only informative about electronic coherences of order $s\le r$.} That is, it cannot distinguish between a superposition and a mixed state between Slater determinants that differ by $r+1$ (or more) particle transitions.  This is in contrast with the $N$-body purity where all possible coherences in the system are evident. 

To make this observation evident, consider the $r$-body  density matrix associated with  the general $N$-particle density matrix  in \eq{eq:rho}, 
\be
{}^{(r)}\Gamma^{j_1 j_2 \cdots j_r}_{i_1 i_2 \cdots i_r} = \frac{1}{r!} \sum_{n, m}  
a_{nm}  \bra{\Phi_m}
\hat{c}_{i_1}^\dagger \hat{c}_{i_2}^\dagger\cdots \hat{c}_{i_r}^\dagger \hat{c}_{j_r} \cdots \hat{c}_{j_2} \hat{c}_{j_1} 
\ket{\Phi_n}.
\ee
The coherences between states $n$ and $m$ in the $N$-particle density matrix can only contribute to the $r$-body reduced density matrix if  $\bra{\Phi_m}
\hat{c}_{i_1}^\dagger \hat{c}_{i_2}^\dagger\cdots \hat{c}_{i_r}^\dagger \hat{c}_{j_r} \cdots \hat{c}_{j_2} \hat{c}_{j_1} 
\ket{\Phi_n}\ne 0$. For this to happen, there has to be some $r$-body  transition that connects the two states. Hence, if the two states differ by $s>r$ particle transitions any coherences that may exist between them is simply not reflected in the $r$-body density matrix and hence in the $r$-body purity. 

\subsection{A closer look into the one-body and two-body  purities} 
\label{stn:p1andp2}

To isolate additional properties of the reduced purities and to illustrate their interpretation, we now determine explicit expressions for $P_1$ and $P_2$  for the general electronic density matrix in \eq{eq:rho}. While it is possible to calculate higher order reduced purities through judicious application of Wick's theorem~\cite{fetter, schaefer2007}, the one-body and two-body purities are the most important and readily applicable cases. To proceed, it is useful to first determine the purity for the simpler case where the density matrix only involves two  $N$-particle Slater determinants 
\be
\label{eq:model}
\hat{\rho}_e = \sum_{n, m=1}^{2} a_{nm} \ket{\Phi_n}\bra{\Phi_m}
\ee
and then extend the solution to an arbitrary number of $\ket{\Phi_n}$ states. Without loss of generality,  we suppose that $\ket{\Phi_2}$ is at most two-particle transitions away from $\ket{\Phi_1}$ since only coherences of order 2 or less are visible in $P_2$.  We  choose $\ket{\Phi_1}$ as the reference state and write,
\be
\label{eq:phi2}
\ket{\Phi_2} = \hat{c}_{\alpha_2}^\dagger \hat{c}_{\beta_2}  \hat{c}_{\alpha_1}^\dagger \hat{c}_{\beta_1} \ket{\Phi_1}.
\ee
In order to guarantee that $\ket{\Phi_1}\ne \ket{\Phi_2}$, we choose $\alpha_1\ne \beta_1$ and $\alpha_2\ne \beta_2$. Since we are interested in $\ket{\Phi_2}\ne 0$, then $\beta_1\ne \beta_2$, $\alpha_1\ne \alpha_2$, and $\hat{c}_{\beta_1}^\dagger \ket{\Phi_1}= \hat{c}_{\alpha_1} \ket{\Phi_1} = \hat{c}_{\alpha_2} \ket{\Phi_1} = 0$.  The particular case where $\ket{\Phi_2}$ and $\ket{\Phi_1}$ differ by a single-particle transition is obtained when $\beta_2 = \alpha_1$. 

From \eq{eq:rpurity}, $P_1$ is given by:
\be
\label{eq:onepurity}
P_1 = \sum_{\epsilon_1, \epsilon_2}
{}^{(1)}{\Gamma}_{\epsilon_1}^{\epsilon_2}
{}^{(1)}{\Gamma}_{\epsilon_2}^{\epsilon_1},
\ee
where, for convenience, the trace has been expressed in the $\epsilon$-basis where  \eq{eq:distrib} holds. In this basis, the one-body  reduced density matrix  for the model state in \eq{eq:model}  is given by
\be
\label{eq:onerdm}
\begin{split}
{}^{(1)}{\Gamma}_{\epsilon_1}^{\epsilon_2}  = & \delta_{\epsilon_1, \epsilon_2} \left[a_{11} f_1(\epsilon_1) + a_{22} f_2(\epsilon_1)  \right] \\
 + & \delta_{\alpha_1, \beta_2} f_1(\beta_1) (1-f_1(\alpha_1))(1-f_1(\alpha_2)) \times\\
 &\left[a_{12}\delta_{\epsilon_1, \alpha_2} \delta_{\epsilon_2, \beta_1} + a_{12}^\star \delta_{\epsilon_1, \beta_1} \delta_{\epsilon_2, \alpha_2}\right],
\end{split}
\ee
where we have taken \eq{eq:distrib} and  \eqref{eq:phi2}  into account and used the fermionic anticommutation relations. Inserting \eq{eq:onerdm} into  \eqref{eq:onepurity} yields
\begin{equation*}
\begin{split}
P_1   = & \sum_\epsilon (a_{11} f_1(\epsilon)+ a_{22} f_2(\epsilon))^2  \\
 + & 2|a_{12}|^2  f_1(\beta_1) ( 1-f_1(\alpha_1))( 1-f_1(\alpha_2)) \delta_{\alpha_1, \beta_2}.
\end{split}
\end{equation*}
Now, supposing that $\ket{\Phi_2}\ne 0$ (such that $ f_1(\beta_1) ( 1-f_1(\alpha_1))( 1-f_1(\alpha_2)) = 1$) and noting that the requirement that $\alpha_1=\beta_2$ is equivalent to requiring $s_{12}=1$ then
\be
\label{eq:modelp1}
P_1   = \sum_\epsilon (a_{11} f_1(\epsilon)+ a_{22} f_2(\epsilon))^2 
 + 2|a_{12}|^2  \delta_{s_{12}, 1},
\ee
which determines the one-body purity for the two-state system in \eq{eq:model}. 

Extending the previous discussion to the general case, the one-body purity for a  many-body state of the form in \eq{eq:rho} is given by
\be
\label{eq:onepuritygrl}
\begin{split}
P_1   & = \sum_\epsilon \big(\sum_{n}a_{nn} f_n(\epsilon)\big)^2 
 + 2\sum_{n>m} |a_{nm}|^2  \delta_{s_{nm}, 1} \\
  & = N - 2\sum_{n>m} \big(a_{nn}a_{mm} s_{nm} - |a_{nm}|^2 \delta_{s_{nm}, 1}\big),
\end{split}
\ee
where the first two terms in the second line depend on the populations in the expansion  of $\hat{\rho}_e$ in~\eq{eq:rho}, while the last one characterizes the contributions due to the coherences. 
In writing \eq{eq:onepuritygrl}, we have extended the process that lead to \eq{eq:modelp1}  to accommodate an arbitrary number of states and taken into account \eq{eq:order}, the  state normalization  $\sum_n a_{nn} = 1$ and the fact that $\sum_\epsilon f_n(\epsilon) = N$.  Note that, as pointed out previously,  $P_1(t)$  decreases with the decoherence between states that differ by one-particle transitions and is unaffected by decoherence processes that involve higher-order coherences.

The derivation of $P_2$ proceeds along similar lines.  In the $\epsilon$-basis,  $P_2$ can be expressed as
\be
\label{eq:2purity}
P_2 = \sum_{\epsilon_1, \epsilon_2, \epsilon_3, \epsilon_4}
{}^{(2)}{\Gamma}_{\epsilon_1, \epsilon_2}^{\epsilon_4,  \epsilon_3}
{}^{(2)}{\Gamma}_{\epsilon_4, \epsilon_3}^{\epsilon_1, \epsilon_2}.
\ee
Here the matrix elements determining $^{(2)}\Gamma_{\epsilon_1, \epsilon_2}^{\epsilon_4,  \epsilon_3}$ and $P_2$ for the two-state model in \eq{eq:model} are given by:
\begin{gather*}
\bra{\Phi_n} \hat{A}  \ket{\Phi_n}  =  f_n(\epsilon_3)  f_n(\epsilon_4) (\delta_{\epsilon_1, \epsilon_4}\delta_{\epsilon_2, \epsilon_3} -\delta_{\epsilon_1, \epsilon_3}\delta_{\epsilon_2, \epsilon_4}) \\
\begin{split}
 \bra{\Phi_1} \hat{A} \ket{\Phi_2} 
 =  f_1(\epsilon_1)  f_1(\epsilon_2)f_1(\beta_1)(1-f_1(\alpha_1))(1-f_1(\alpha_2))[   \\
  \delta_{\alpha_1, \beta_2}( 
\delta_{\epsilon_1, \beta_1} (\delta_{\epsilon_2, \epsilon_3} \delta_{\epsilon_4, \alpha_2} - \delta_{\epsilon_2, \epsilon_4} \delta_{\epsilon_3, \alpha_2}) \\
- 
\delta_{\epsilon_2, \beta_1}(\delta_{\epsilon_1, \epsilon_3} \delta_{\epsilon_4, \alpha_2} - \delta_{\epsilon_1, \epsilon_4} \delta_{\epsilon_3, \alpha_2})) \\
+  (\delta_{\epsilon_2, \beta_1} \delta_{\epsilon_1, \beta_2} - \delta_{\epsilon_2, \beta_2} \delta_{\epsilon_1, \beta_1})
(\delta_{\epsilon_4, \alpha_2}  \delta_{\epsilon_3, \alpha_1} - \delta_{\epsilon_4, \alpha_1} \delta_{\epsilon_3, \alpha_2})
]
\end{split}\\
\begin{split}
\bra{\Phi_2} \hat{A}\ket{\Phi_1} 
 =  f_1(\epsilon_3)  f_1(\epsilon_4)f_1(\beta_1)(1-f_1(\alpha_1))(1-f_1(\alpha_2))[  \\
  \delta_{\alpha_1, \beta_2}(
\delta_{\epsilon_4, \beta_1} (\delta_{\epsilon_2, \epsilon_3} \delta_{\epsilon_1, \alpha_2} - \delta_{\epsilon_1, \epsilon_3} \delta_{\epsilon_2, \alpha_2}) \\
- \delta_{\epsilon_3, \beta_1}(\delta_{\epsilon_2, \epsilon_4} \delta_{\epsilon_1, \alpha_2} - \delta_{\epsilon_1, \epsilon_4} \delta_{\epsilon_2, \alpha_2})) \\
+  (\delta_{\epsilon_3, \beta_1} \delta_{\epsilon_4, \beta_2} - \delta_{\epsilon_3, \beta_2} \delta_{\epsilon_4, \beta_1})
(\delta_{\epsilon_1, \alpha_2} \delta_{\epsilon_2, \alpha_1} - \delta_{\epsilon_1, \alpha_1} \delta_{\epsilon_2, \alpha_2})
],
\end{split}
\end{gather*}
where $\hat{A} \equiv \hat{c}_{\epsilon_1}^\dagger \hat{c}_{\epsilon_2}^\dagger \hat{c}_{\epsilon_3} \hat{c}_{\epsilon_4} $. Using these matrix elements,  it follows that 
\begin{equation*}
\begin{split}
P_2 
& =\frac{1}{4} \sum_{\epsilon_1, \epsilon_2, \epsilon_3, \epsilon_4}   \textrm{Tr}\{\hat{A} \hat{\rho}_e \}\textrm{Tr}\{\hat{A}^\dagger  \hat{\rho}_e\} \\
 & = \frac{1}{4} \sum_{\epsilon_1, \epsilon_2, \epsilon_3, \epsilon_4}   \Big[
 \sum_{n, m=1}^{2} a_{nn} a_{mm}   \bra{\Phi_n}\hat{A} \ket{\Phi_n} \bra{\Phi_m}\hat{A} ^\dagger\ket{\Phi_m}\\
& +  |a_{12}|^2 \big( \bra{\Phi_1} \hat{A} \ket{\Phi_2}\bra{\Phi_2}  \hat{A}^\dagger \ket{\Phi_1}  
+ \bra{\Phi_2} \hat{A}\ket{\Phi_1}\bra{\Phi_1} \hat{A}^\dagger  \ket{\Phi_2}
\big)\Big],
\end{split}
\end{equation*}
where all other terms in the product are zero. Calculating explicitly each of the remaining terms, the two-body reduced purity for the model density matrix is given by: 
\begin{equation*}
\begin{split}
P_2   =\sum_{n, m=1}^{2} &  \frac{a_{nn} a_{mm}}{2} 
\big[ \big(\sum_\epsilon f_n(\epsilon) f_m(\epsilon) \big)^2 - \sum_\epsilon f_n(\epsilon) f_m(\epsilon) \big] \\
 +& 2|a_{12}|^2  f_1(\beta_1)(1-f_1(\alpha_1))(1-f_1(\alpha_2)) \times\\
 &\left[\delta_{\alpha_1, \beta_2} (N-1) + f_1(\beta_2)\right].
\end{split}
\end{equation*}
This expression can be cast into a form that is simpler to generalize by taking into account  that if $\ket{\Phi_2}\ne 0$, $f_1(\beta_1)(1-f_1(\alpha_1))(1-f_1(\alpha_2))=1$; and that when $\alpha_1=\beta_2$ (or  $f_1(\beta_2)=1$) the order of the coherence is $s_{12}=1$ (or $s_{12} =2$).  Thus, 
\begin{equation*}
\begin{split}
P_2  = \sum_{n, m=1}^{2} & \frac{a_{nn} a_{mm}}{2} \big[ \big(\sum_\epsilon f_n(\epsilon) f_m(\epsilon) \big)^2 - \sum_\epsilon f_n(\epsilon) f_m(\epsilon)\big] \\
& + 2|a_{12}|^2  \left[\delta_{s_{12}, 1} (N-1) + \delta_{s_{12},2}\right].
\end{split}
\end{equation*}
This expression can be extended to capture the behavior of the general many-body state in \eq{eq:rho} by taking into account the contribution of all possible pairs of states. In this case,
\be
\label{eq:twopuritygrl}
\begin{split}
P_2  & =  \sum_{n, m} \frac{a_{nn} a_{mm} }{2}
\Big[ \big(\sum_\epsilon f_n(\epsilon) f_m(\epsilon) \big)^2 - \sum_\epsilon f_n(\epsilon) f_m(\epsilon) \Big] \\
& + 2\sum_{n>m}|a_{nm}|^2  \left[\delta_{s_{nm}, 1} (N-1) + \delta_{s_{nm},2}\right]\\
 & = \frac{N(N-1)}{2}  -  \sum_{n>m} a_{nn} a_{mm} s_{nm}(2N - s_{nm}-1) \\
 & +  \sum_{n>m} 2|a_{nm}|^2 (\delta_{s_{nm},1}(N-1) + \delta_{s_{nm,2}}),
\end{split}
\ee
where we have used \eq{eq:order} and the fact that $\sum_n f_n(\epsilon) = N$.  The first two terms are due to the populations, while the last term is due to the coherences among the Slater determinants. Note that $P_2$ decays with the loss of coherences of order 1 and 2.

Equations~\eqref{eq:onepuritygrl} and \eqref{eq:twopuritygrl} exemplify the behavior of the one-body and two-body purities for a general electronic state.  In deriving these equations we have taken advantage of the structure of Slater determinants. Note, however, that the value of the reduced purities is representation-independent and does not change if a different complete single-particle basis $\{\ket{\epsilon}\}$ is employed or if no decomposition into Slater determinants is invoked.  This is evidenced by \eq{eq:purityeigen} that shows the relationship between the reduced purities and the eigenvalues of the reduced density matrices ${}^{(r)} \lambda_n$; the ${}^{(r)} \lambda_n$ are representation independent and hence the reduced purities are also representation independent. This allows for the interpretation of the decoherence in a particular basis without loss of generality.

 We now discuss a few  observations that follow from these general expressions.

\subsection{Electronic correlation and the interpretation of the reduced purities} 
\label{stn:interpretation}

The reduced purities offer a window into the coherence behavior of many-electron systems and allow isolating coherence effects of a particular order. Nevertheless, because the $P_r$ are defined from reduced electronic quantities, their interpretation can be more challenging than the one of the $N$-body purity.  Note, in particular,  that while the value of the $N$-body purity for a pure state is always 1, the value of $P_r$ for pure electronic states depends on the degree of electronic correlation.  Electronic correlation leads to nonidempotency in the reduced density matrices (see, e.g., Refs.~\onlinecite{ziesche95, ziesche97}) and thus to a reduction in the reduced purity that is not due to bath-induced decoherence. As a consequence, an observed decay in the reduced purity can be due to a decay in the coherence properties of the system, or due to a change in the correlations of the many-electron system even in the absence of decoherence. 

Note that it is technically possible to construct electronic decoherence measures based on the reduced purities that solely reflect decoherence processes. To see this, it is useful to recall the  Carlson-Keller theorem~\cite{carlsonkeller} which states that for pure bound states the nonzero eigenvalues of ${}^{(r)}\hat{\Gamma}$ and  ${}^{(N-r)}\hat{\Gamma}$ are identical. Since the reduced purities are determined by such eigenvalues [recall \eq{eq:purityeigen}], then a quantity like $P_{N-r}- P_r$ would be identically zero for pure states and nonzero for mixed states, irrespective of the details of the system-bath evolution.  While of formal interest, such measures of electronic decoherence are of little practical use because they require knowledge of high-order density matrices that are generally not available. 

\begin{figure}[htbp]
\centering
\includegraphics[width=0.45\textwidth]{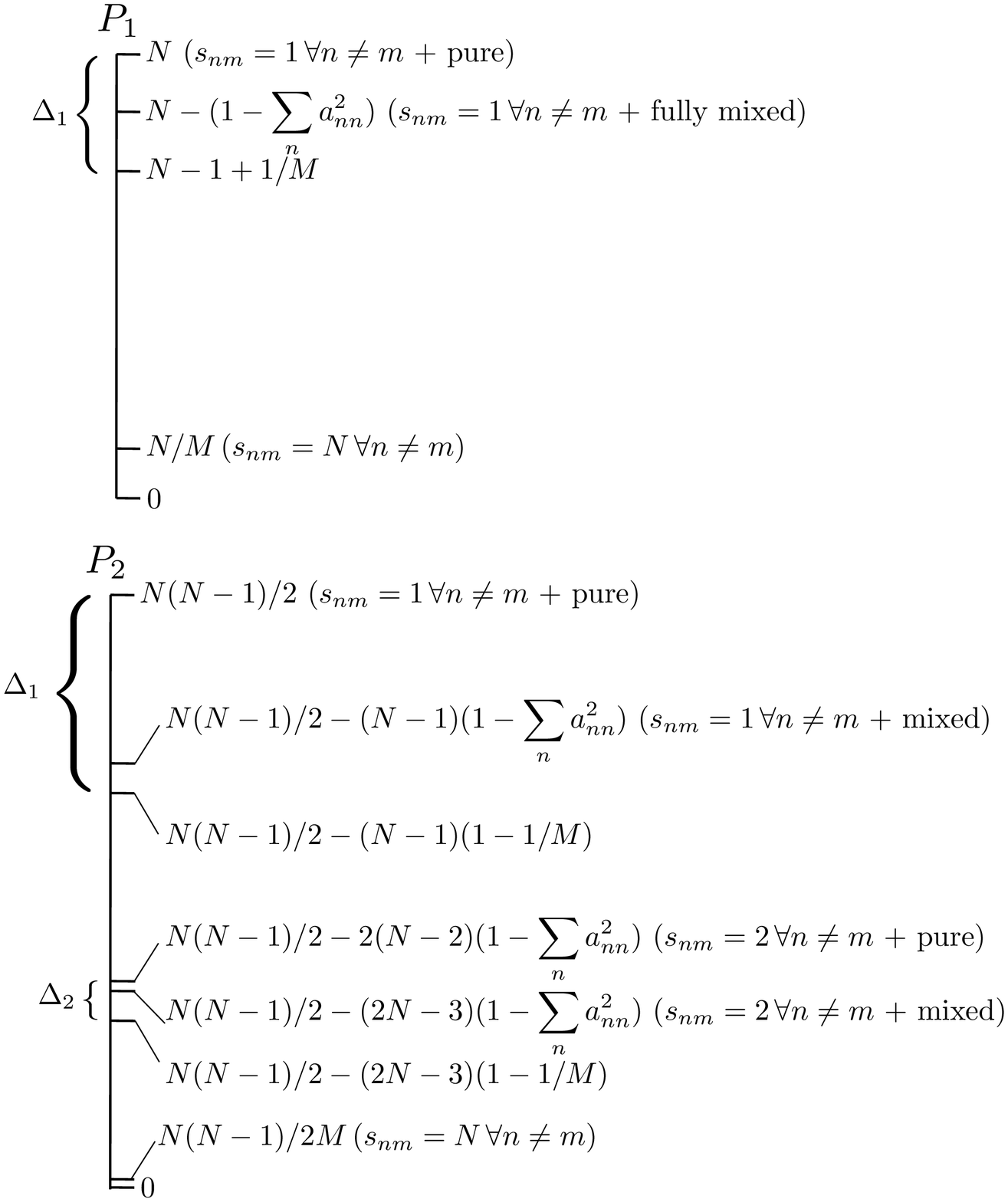}
\caption{Limiting values for the one-body $P_1$ [\eq{eq:onepuritygrl}] and two-body $P_2$ [\eq{eq:twopuritygrl}]  purities (see text). The quantities $\Delta_1$ and $\Delta_2$ are the maximum possible decay of the reduced purities due to one-body and two-body decoherence, $N$ refers to the number of electrons and $M$ to the number of Slater determinants involved.  The figure is not to scale. }
     \label{fig:puritydiags}
\end{figure}

To understand further the structure and the information that can be gleaned from the reduced purities, consider now the limiting behavior of $P_1$ and $P_2$. For reference in the discussion we have tabulated the main limiting values of $P_1$ and $P_2$ in Figure~\ref{fig:puritydiags}. From \eq{eq:onepuritygrl}, the maximum value for $P_1$ is $N$, obtained when only a single Slater determinant is involved or when all terms in the superposition are such that $s_{nm}=1$ and  $|a_{nm}|^2 = a_{nn}a_{mm}$. In the absence of population changes, a decay in $P_1$ follows the  decay of one-body coherences; the decoherence between states $n$ and $m$ with $s_{nm}=1$ can induce a maximum decay of $2|a_{nm}|^2$.  
Given a set of populations $\{ a_{nn} \}$, the minimum value that $P_1$ can achieve solely due  to decoherence is  $P_1 = N -1 + \sum_n a_{nn}^2$, obtained when $s_{nm}=1$ and $a_{nm}= 0$ for all pairs of states. Thus, the maximum possible decay in the one-body reduced purity due to decoherence is $\Delta_1= 1-1/M$  and occurs when all $M$ Slater determinants are equally populated and maximally entangled with the bath.  As a consequence of this, a decay of the one-body purity beyond $\Delta_1$ cannot be explained solely on the basis of decoherence of first order and is indicative of the involvement of states with $s_{nm}$'s of higher order. In fact, the absolute minimum of $P_1$ occurs when the density matrix is composed of equally populated states $a_{nn}=1/M$ that all differ by $N$-particle transitions among each other (i.e. $s_{nm}=N, \, \forall n\ne m$). In this case $P_1 = N/M$ irrespective of whether the state is a superposition state or an incoherent mixture. 

The limiting cases for $P_2$ are shown in the lower panel of \fig{fig:puritydiags}. The maximum value of the two-body purity [\eq{eq:twopuritygrl}] is $P_2=N(N-1)/2$ obtained for a single Slater determinant or for a coherent superposition where  $s_{nm}=1,\, \forall n\ne m$. In turn, the  minimum value of $P_2= N(N-1)/(2M)$ is  obtained when all $M$ participating Slater determinants are equally populated ($a_{nn}=1/M$) and differ by $N$-particle transitions (i.e., $s_{nm}=N$), irrespective of whether the $N$-body density matrix represents a pure state or not. In the absence of  changes in the $a_{nn}$'s, a decay in $P_2$ signals coherence loss of order 1 or 2. Importantly, note that the magnitude of the decay actually depends on the order of the coherence that is lost; the lowest order coherences having the highest impact on the reduced purity.  Specifically, the decoherence of a superposition of states differing by single-particle transitions leads to a decay of $P_2$ from  $P_2=N(N-1)/2$ to  $P_2 = N(N-1)/2 - (N-1)(1- \sum_n a_{nn}^2)$, for a maximum decay of $\Delta_1=(N-1)(1- 1/M)$.  In turn, the decoherence of a superposition of states that differ by two-particle transitions leads to a reduction from $P_2=  N(N-1)/2 - 2(N-2)(1- \sum_n a_{nn}^2)$ to $P_2=N(N-1)/2 - (2N-3)(1- \sum_n a_{nn}^2)$, for a maximum decay  of $\Delta_2=(1-1/M)$ which is $(N-1)$ times less than the reduction $\Delta_1$ due to decoherence between states for which $s_{nm}=1$. A value of $P_2$ lower than $P_2= N(N-1)/2 - (2N-3)$ necessarily indicates that there are $s_{nm}>2$ in the states involved.

As seen in Eqs.~\eqref{eq:onepuritygrl} and \eqref{eq:twopuritygrl}, the decay of the reduced purities directly signals coherence loss in  ``pure dephasing" cases where the system-bath evolution does not lead to appreciable changes in the populations of the Slater determinants involved. More generally, the populations of the Slater determinants can change in time due to interactions of the electrons with themselves, with bath degrees of freedom or with an external potential (i.e. a laser). In such general case,  in order to cleanly identify the decoherence contributions to  the dynamics of the reduced purities it is required to know the populations and the distribution functions of the Slater determinants  involved. 
This feature is the main limiting factor in the utility of the reduced purities in characterizing decoherence effects for, generally, from a reduced density matrix it is not  always easy to uniquely  unravel  the populations of the underlying  possible Slater determinants used to describe the correlated electronic states.  

However,  if additional details of the problem are known, like the active determinant space and the initial state,  it then becomes increasingly plausible to perform a detailed analysis of the reduced purities on the basis of \eq{eq:onepuritygrl} and~\eq{eq:twopuritygrl} even in situations when the populations of the Slater determinants are continuously changing. We now briefly sketch how the coherence properties can be characterized: 1. Specify an active determinant space that is adequate for the problem and identify all possible determinant combinations within this space that are consistent with the orbital populations.  Clearly, a very large active determinant space may make the search intractable, while a too restrictive active space may not lead to a correct characterization. 2.  Fit, in each of those cases, the $a_{nn}$'s to reproduce the  observed orbital population dynamics. If the procedure is not unique, keep track of competing possibilities.  3.  Given each individual set of model $\{a_{nn}\}$, using \eq{eq:onepuritygrl} calculate two limits for the one body purity;  a fully incoherent limit $P_{1}^{\textrm{(inc)}}$ where the $a_{nm}=0$ for $n\ne m$ and a coherent limit $P_{1}^{\textrm{(coh)}}$ where  $a_{nn}a_{mm}= |a_{nm}|^2$. 4. Use the observed $P_1$ to discard  possibilities. If $P_{1}^{\textrm{(inc)}}> P_1$ or $P_{1}^{\textrm{(coh)}}< P_1$ discard the possibility, as the model state cannot possibly describe the system. 5. If the coherence properties of the initial state are known, further discard options by demanding the model to exactly reproduce $P_1(0)$.  6. If the procedure did not yield a unique choice, then it is necessary to examine the two-body purity. Repeat steps 3-5 taking advantage of \eq{eq:twopuritygrl}. If this is not enough to yield a unique choice, then the procedure needs to be repeated for increasingly higher order purities until all available information has been exhausted or a unique choice has been determined. Note that coherences of higher order than the highest order purity available would not be able to be resolved.   Section~\ref{stn:examples} discusses  representative examples of such a reconstruction.

\section{Some examples}
\label{stn:examples}

We now illustrate the use of the reduced purities using the example of electronic decoherence in a molecular system due to  electron-vibrational couplings. Specifically, we  consider an oligoacetylene chain with 4 carbon atoms and 4 $\pi$ electrons as described by the Su-Schrieffer-Heeger (SSH) Hamiltonian~\cite{SSH}, a tight-binding model  with electron-vibrational interactions. The details of the Hamiltonian and the Ehrenfest mixed quantum-classical method employed to follow the  vibronic dynamics have been specified before~\cite{franco08, franco12, francoNJP2013}.  What is of relevance to this discussion is that the system consists of 4 noninteracting $\pi$  electrons distributed among 4 spectrally isolated molecular orbitals $\ket{\epsilon_n}$ of energy $\epsilon_n$ that allow for double occupancy, for a total of 19 possible $N$-particle levels (without counting spin-degeneracies) subject to decoherence. The orbital energy and labels in the ground state optimal geometry of the chain are shown in \fig{fig:puritymodelsuper}. Computationally, we  follow the dynamics of the one-body and two-body reduced density matrix for this system and use it to determine $P_1(t)$ and $P_2(t)$.

\subsection{Decoherence of model superposition states}
\label{stn:modelsuperposition}

Consider first a ``pure dephasing" example where there are no changes in the population of the Slater determinants during the dynamics. In this case, the decay of the reduced purities are directly indicative of decoherence. Specifically, we follow the system-bath dynamics after preparation of the composite system in an initial separable superposition state of the form:
\be
\label{eq:initsuper}
\ket{\Omega(0)} = ( {c}_0\ket{\Phi_0} + {c}_1 \ket{\Phi_1} )\otimes \ket{\chi_0},\\
\ee
where $\ket{\chi_0}$ is the ground vibrational state in the ground electronic state $\ket{\Phi_0}$, and $\ket{\Phi_1}$ is an excited state. The $\ket{\Phi_1}$ is selected to be spectrally isolated from other $N$-particle states such that the vibronic evolution does not lead to population exchange into other levels, as revealed by constant orbital populations throughout the dynamics.  

\begin{figure}[htbp]
\centering
\includegraphics[width=0.45\textwidth]{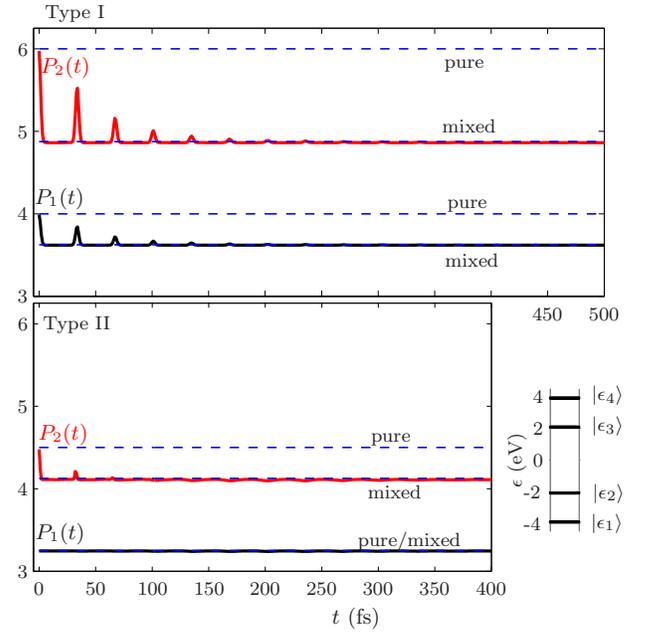}
\caption{Reduced purities during the vibronic evolution of a neutral oligoacetylene SSH chain with $N=4$ electrons. The system is initially prepared in a superposition $\ket{\Psi(0)} = ( {c}_0\ket{\Phi_0} + {c}_1 \ket{\Phi_1} )\otimes \ket{\chi_0}$ ($|{c}_0|^2=3/4$ and $|{c}_1|^2=1/4$) between the ground state $\ket{\Phi_0}$ and an excited electronic state $\ket{\Phi_1}$. The initial nuclear state $ \ket{\chi_0}$ is chosen to be the ground vibrational state associated with $\ket{\Phi_0}$. Type I: First order coherence, $\ket{\Phi_1}=\hat{c}_{\epsilon_3\uparrow}^\dagger  \hat{c}_{\epsilon_2\uparrow} \ket{\Phi_0}$. Type II: Second order coherence, $\ket{\Phi_1}=   \hat{c}_{\epsilon_3 \downarrow}^\dagger \hat{c}_{\epsilon_2\downarrow} \hat{c}_{\epsilon_3\uparrow}^\dagger \hat{c}_{\epsilon_2\uparrow} \ket{\Phi_0}$. The dashed lines signal the fully coherent/incoherent limit of $P_1$ and $P_2$ computed using Eqs. \eqref{eq:onepuritygrl} and \eqref{eq:twopuritygrl} assuming that only $\ket{\Phi_0}$ and $\ket{\Phi_1}$ participate in the dynamics.  The orbital labels and  energies at initial time are shown in the bottom-right corner. }
     \label{fig:puritymodelsuper}
\end{figure}

Two different types of initial superposition states are considered. In type I, $\ket{\Phi_1}=\hat{c}_{\epsilon_3\uparrow}^\dagger  \hat{c}_{\epsilon_2\uparrow} \ket{\Phi_0}$ is obtained from the ground state via a HOMO$\to$LUMO transition in a given spin channel, and the coherence order is 1. In type II, $\ket{\Phi_1}=   \hat{c}_{\epsilon_3, \downarrow}^\dagger \hat{c}_{\epsilon_2, \downarrow} \hat{c}_{\epsilon_3, \uparrow}^\dagger \hat{c}_{\epsilon_2, \uparrow} \ket{\Phi_0}$ is a doubly excited state where the two electrons in the HOMO of $\ket{\Phi_0}$ are promoted into the LUMO, and the resulting coherence is of second order.   Figure \ref{fig:puritymodelsuper} shows the dynamics of the purities in these two cases for $|{c}_0|^2=3/4$ and $|{c}_1|^2=1/4$. The dashed lines in the figure indicate the fully coherent/incoherent behavior expected for $P_1$ and $P_2$ as computed from Eqs. \eqref{eq:onepuritygrl} and \eqref{eq:twopuritygrl}  assuming that only $\ket{\Phi_0}$ and $\ket{\Phi_1}$   participate in the dynamics. In interpreting the results, it is useful to keep \fig{fig:puritydiags} in mind.

Focus first on the dynamics of the type I superposition (\fig{fig:puritymodelsuper}, top panel).   At initial time $P_1=N$ and $P_2= N(N-1)/2$ because the system is pure and the coherence is of first order. The system-bath evolution leads to a decay of the purities that is entirely due to decoherence. Since $s_{10}=1$, both $P_1$ and $P_2$ follow the coherence decay, and the fall of $P_2$ is $(N-1)$ times larger than the one of $P_1$. The partial recurrences in the purities signal  vibronic evolution of the chain~\cite{franco12}.  After 200 fs the system is well described as an incoherent mixture.  In the type II case (\fig{fig:puritymodelsuper}, bottom panel), $P_2$ follows the decoherence while $P_1$ remains constant because it cannot distinguish  a coherence of second order from a mixture of states. At initial time, $P_2$ takes its maximum value that is consistent with the superposition in question and evolves with the vibronic evolution. The dynamics of $P_2$ cleary shows decoherence in $\sim100$ fs of a superposition of second order. Note that the observed decay of  $P_2$ in this case is quantitatively smaller than the one observed in a first order coherence since the decoherence of lowest order has a larger impact in the two-body purity (recall \fig{fig:puritydiags}).

\subsection{Resonant photoexcitation}

\begin{figure}[htbp]
\centering
\includegraphics[width=0.45\textwidth]{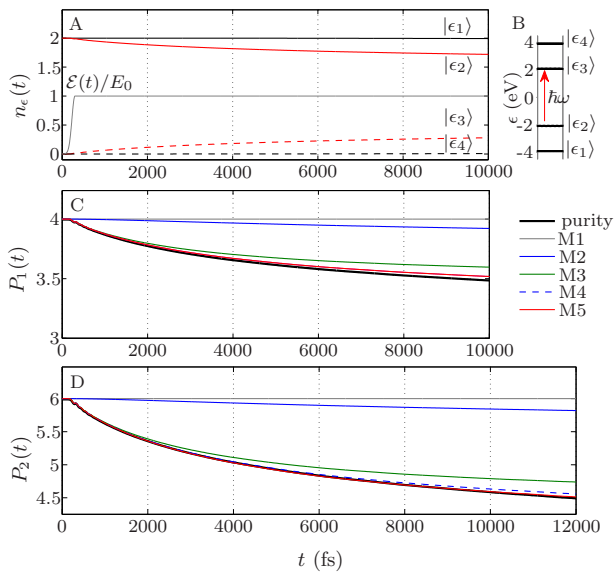}
\caption{Reduced purities during photoexcitation of a neutral SSH chain with 4 electrons. (\textbf{A}) Orbital populations during  photoexcitation with a laser field $E(t) = \mc{E}(t) \cos(\w t)$. The laser  frequency $\hbar \w=4.08$ eV   is chosen to be at resonance with the HOMO-LUMO transition.  The envelope function $\mc{E}(t)$ has a Gaussian turn-on and remains constant at $E_0=8.7\times 10^{-3}$ V/\AA~ from $t=300$ fs on. (\textbf{B}) Single-particle spectrum of the SSH chain at initial time. (\textbf{C})  One-body and (\textbf{D})  two-body purity during the dynamics. The black lines  are the result of an explicit simulation of the vibronic dynamics. The colored lines correspond to different models with varying degrees of coherence assumed in the  $N$-body density matrix (see text). In (C) the lines for \textbf{M3} and \textbf{M4} are on top of each other and cannot be distinguished. }
     \label{fig:purity}
\end{figure}

To illustrate the use of the reduced purities in a more complex  setting, we now consider electronic decoherence due to vibronic interactions  during resonant photoexcitation of a molecular  system.  This example illustrates how through an analysis of the reduced purities it is possible to establish the coherence properties of an $N$-particle system even when only the reduced density matrices are known. Contrary to the previous example, because of the photoexcitation, the population of the involved Slater determinants changes continuously during the dynamics.

Figure~\ref{fig:purity} shows the orbital energies, orbital populations and  reduced purities of the SSH chain during dipole-interaction  with a continuous wave  laser that is resonant with the HOMO-LUMO transition.  Initially, the system is prepared in the ground vibronic state with an electronic state $\ket{\Phi_0}$ where the lowest energy molecular orbitals $\ket{\epsilon_1}$ and $\ket{\epsilon_2}$ are doubly occupied. As shown in \fig{fig:purity}A-B, the laser promotes population into the $\ket{\epsilon_3}$ orbital. The reduced purities resulting from the numerical simulation are shown in black in \fig{fig:purity}C-D. Because the laser induces changes in the state populations, simple inspection of $P_1$ and $P_2$ cannot reveal the nature of the coherences and an explicit model of the dynamics is required.   From an $N$-particle perspective,  the laser field can transfer population from $\ket{\Phi_0}$ into the degenerate pair
\be
\ket{\Phi_1} = \hat{c}_{\epsilon_3\uparrow}\hat{c}_{\epsilon_2\uparrow}\ket{G}, \quad  
\ket{\Phi_2} = \hat{c}_{\epsilon_3\downarrow}\hat{c}_{\epsilon_2\downarrow}\ket{G}.
\ee
Supposing that only states $\ket{\Phi_0}$, $\ket{\Phi_1}$ and $\ket{\Phi_2}$ can participate in the dynamics, the $N$-particle density matrix of the system can be written as:
\be
\label{eq:mdm}
\hat{\rho}_e = \sum_{n=0}^2\sum_{m=0}^2 a_{nm}  \ket{\Phi_n}\bra{\Phi_m}.
\ee
We consider  different models for $\hat{\rho}_e$ based on \eq{eq:mdm}  that differ in the degree of assumed coherence and the states involved. Specifically, we consider models
\begin{align*}
\label{eq:models}
&\textbf{M1:} \quad  &|a_{nm}|^2  &=a_{nn} a_{mm};   &a_{22}&= 0; \\
&\textbf{M2:}\quad  &|a_{nm}|^2 & =a_{nn} a_{mm};  &a_{11}&=a_{22};\\
&\textbf{M3:}\quad    &a_{2n} & =0, \, \forall n;  &a_{01}&= 0; \\
&\textbf{M4:} \quad   &a_{01} & =a_{02}=0; & a_{11}&=a_{22}; \\
&\textbf{M5:}  \quad  &a_{nm} & = 0,\, \forall n\ne m;  & a_{11}&=a_{22}.   
\end{align*}
These models correspond, respectively, to  (\textbf{M1}) a coherent superposition  between states $\ket{\Phi_0}$ and $\ket{\Phi_1}$ with no participation of $\ket{\Phi_2}$; (\textbf{M2}) a coherent superposition involving all three states;  (\textbf{M3}) an incoherent mixture between states $\ket{\Phi_0}$ and $\ket{\Phi_1}$; (\textbf{M4}) a partially coherent triad where only the coherences between $\ket{\Phi_1}$ and $\ket{\Phi_2}$ are maintained; and (\textbf{M5}) a fully incoherent state. In the models where all three states are considered (M2, M4 and M5)  we take $a_{11}=a_{22}$ since the $\ket{\Phi_0}\to\ket{\Phi_1}$ and $\ket{\Phi_0}\to\ket{\Phi_2}$ transition dipoles are identical. 
As shown below, only when the correct form for $\hat{\rho}_e$ is assumed the reduced purities reconstructed from  Eqs.~\eqref{eq:onepuritygrl} and \eqref{eq:twopuritygrl} match the reduced purity obtained directly from the numerical simulation.

In \textbf{M1},  $P_1=N$ and $P_2=N(N-1)/2$  (\fig{fig:purity}, grey lines) since   two states involved are separated by a single-particle transition and are coherent  (recall \eq{eq:onepuritygrl} and~\eq{eq:twopuritygrl}). This model fails to capture the observed decay of the purities as the photoexcitation proceeds and is not a faithful description of the system. 
The reduced purities for  \textbf{M2} observe a decay (blue lines in \fig{fig:purity}) that is not due to decoherence but  due to the transfer of population into a pair of states $\ket{\Phi_1}$ and $\ket{\Phi_2}$ with $s_{12}=2$. Nevertheless, the decay in the reduced purities in this model does not capture that observed in the simulation, indicating that a fully coherent model is not a faithful description of the system. In \textbf{M3}, the assumed decoherence between states $\ket{\Phi_0}$ and $\ket{\Phi_1}$  leads to a significant decrease in $P_1$ and $P_2$ (green lines, \fig{fig:purity}). However, the decay is not sufficient to explain the observed behavior indicating that a model where all 3 states are taken into account is required. Both three-state models \textbf{M4} (blue dashed lines) and \textbf{M5} (red lines) reproduce equally well the behavior of $P_1$. In fact, they are indistinguishable in $P_1$ since \textbf{M4}  contains 2nd order coherences not present in the fully incoherent model  \textbf{M5}  that cannot be resolved by $P_1$. In order to determine which state represents best the state of the system an analysis of $P_2$ is required. As shown in \fig{fig:purity}D, the model that best adjusts to the observed behavior  is \textbf{M5} indicating that during photoexcitation  the system is best described as an incoherent mixture between states $\ket{\Phi_0}$, $\ket{\Phi_1}$ and $\ket{\Phi_2}$.  This is because of the fast electronic decoherence timescale that is characteristic of the model and method employed~\cite{franco12, francoNJP2013}.

\section{Conclusions}
\label{stn:conclusions}

A family of reduced purity measures that can be used to characterize decoherence phenomena in many-electron systems has been introduced based on the hierarchy of electronic reduced density matrices. Using the properties of Slater determinants, explicit expressions for the one-body and two-body  purities for a general electronic state have been derived and used to elucidate the structure and information content of the reduced purities. As shown, the reduced purities can be used to characterize electronic decoherence when only  few-body electronic reduced density matrices are known. Further, the measures permit decomposing electronic decoherence phenomena into contributions arising from coherences of different order  providing, in this way, a useful interpretative tool of the dynamics.  The  use of the reduced purities was exemplified via investigation of decoherence in a model molecular system with electron-vibrational interactions  both in a pure dephasing case and in a case where the electronic structure is constantly changing due to resonant photoexcitation. 

Subtleties can develop in the interpretation of the reduced purities because of the fact that we deal with a general many-electron subsystem but only use reduced information about the electronic degrees of freedom.
In particular, a decay in the reduced purities is seen to arise either due to bath-induced decoherence or due to an increase in electronic correlation as both phenomena lead to nonidempotency of the reduced electronic density matrices.   While it is technically clear how to define reduced purity measures that solely reflect decoherence properties via the Carlson-Keller theorem, these measures  require knowledge of higher order electronic density matrices that are typically not available.

In the particular case of pure dephasing problems the interpretation of the reduced purities is straightforward as a decay of the reduced purities directly signals coherence loss. For the more general case, a systematic  procedure to determine the decoherence contributions to the reduced purities was presented. Since such a procedure involves unraveling the observed dynamics of the few-body electronic density matrices into the $N$-particle Hilbert space, isolating a unique solution  can only benefit from any additional information about the electronic subsystem that is available such as the active determinant space and the initial state. 

At this point, it is useful to connect the proposed reduced purity measures with existing electronic structure, condensed matter and quantum optics formalisms. From a time-dependent density functional perspective (TDDFT), in principle the reduced purities are functionals of the time-dependent density and the Kohn-Sham and many-body initial states~\cite{runge1984,kreibich2001,TDDFT2012} even for an open  electronic subsystem. This functional dependence allows expressing 
the off-diagonals of all electronic reduced density matrices in terms of the diagonal elements of the reduced one-body density matrix in position representation. However, to date, this functional dependence is not fully known and cannot be exploited to further advance the present considerations. Other electronic structure theories that employ reduced density matrices, like reduced density matrix functional theory (RDMFT) \cite{coleman1963,gilbert1975,gritsenko2005}, currently focus 
on static problems of closed electronic systems. In these theories, the time-dependence and  
decoherence aspects of the reduced purities are expected to be of future relevance. Last, 
 similar purity measures are also applicable in quantum optics~\cite{mandelwolf1995,scully1997,loudon2000} and condensed matter physics~ \cite{fetter,martinschwinger1959} formalisms that employ the hierarchy of $r$-body Green's functions. This is because the $r$-body density matrices can be obtained from the equal-time limit of $r$-body Green's functions.

While this analysis has focused on purity related measures, the methods, insights and limitations apply to any other measure of decoherence that is based on the density matrix such as the von Neumann entropy.  Future prospects include studying the asymptotic thermal behavior of the reduced purities and the utility of these measures in characterizing increasingly more complex electron-bath dynamics.

\begin{acknowledgments}
I.F. thanks the Alexander von Humboldt Foundation for financial support. The authors thank Prof. Matthias Scheffler for his support during the preparation of this manuscript.
\end{acknowledgments}

%

\end{document}